\documentclass[pra,twocolumn,groupedaddress,floatfix,showpacs,showkeys]{revtex4}

\usepackage[dvips]{graphicx}%
\usepackage{amsmath}
\usepackage{bm}

\begin{document}

\title{Security of Continuous-variable quantum cryptography using coherent states: Decline of postselection advantage}

\author{Ryo Namiki}
\email[Electric address: ]{namiki@qo.phys.gakushuin.ac.jp} 

\affiliation{CREST Research Team for Photonic Quantum Information, Division of Materials Physics, 
Department of Materials Engineering Science, 
Graduate school of Engineering Science, Osaka University, 
Toyonaka, Osaka 560-8531, Japan}

\author{Takuya Hirano }
\affiliation{Department of Physics, Gakushuin University,
Mejiro 1-5-1, Toshima-ku, Tokyo 171-8588, Japan}

\date{\today}
\begin{abstract}
We investigate the security of continuous-variable (CV) quantum key distribution (QKD) using coherent states in the presence of quadrature excess noise.
We consider an eavesdropping attack which uses a linear amplifier and beam splitter. This attack makes a link between beam-splitting attack and intercept-resend attack (classical teleportation attack). 
We also show how postselection loses its efficiency in a realistic channel.   
\end{abstract}

\maketitle


Quantum key distribution (QKD) is a technique that allows two parties, Alice (the sender) and Bob (the receiver), to share a key which is kept secret from an eavesdropper (Eve) who has advanced computational and technological power \cite{rmp74}.
To achieve a signal transmission between distant parties, controlling optical quantum states is essential. Several QKD schemes based on continuous-variable (CV) which uses the quadrature amplitude of light field have been proposed \cite{squeezed,epr,osci,con,alcon,coherent,postsel,coherentR,hirano,namiki1}.
Although usage of squeezed states or EPR states are fundamentally interesting, coherent-state protocols have practical advantage of easy state preparation. CV QKD using coherent states over a 1$k$m-optical-fiber path has been experimentally demonstrated at 1.55$\mu$m-communication wavelength \cite{hirano}.

The performance of QKD is limited by the presence of the transmission loss.
A simple treatment of the loss effect is beam-splitting attack (BSA) where Eve replaces the transmission path with the lossless one and a beam splitter (BS). Then she obtains the signals corresponding to the loss without making any disturbance to the signal.
At first sight, over the existence of 50\% loss (3dB loss), it seems to be impossible to distill the secret key using coherent-state signal because
Eve can get stronger signal than Bob \cite{coherent}. However, since the knowledge about the signal depends on the measurement result, coherent-state protocol can provide a secure key by conditional use of measurement results (postselection, PS) even in the presence of higher loss \cite{postsel,hirano,namiki1}. PS plays an important role in many implementations of quantum information processing tasks as well as QKD.

In realistic condition, besides the loss, excess Gaussian noise is imposed on the quadrature distribution \cite{hirano}.  
Since any excess noise tapers off when the state falls into vacuum at high loss,
the excess noise added by Eve near Alice's side will disappear at Bob's side for a long transmission distance. Then, for a sufficiently long distance, eavesdropping cannot be detectable. From this observation, it is shown that CV-QKD protocols using coherent states cannot work for arbitrary transmission distance in the presence of excess noise \cite{namiki2}. This limitation is given by an intercept-resend attack called classical teleportation attack (CTA).

The question is what kind of attack links between ``direct'' CTA and ``indirect'' BSA, and how PS works in the presence of excess noise. 
In this Letter, we provide an intermediate attack between BSA and CTA, and show how PS loses its advantage in the presence of noise.

We consider the realistic channel which transforms coherent state into a Gaussian mixture of coherent states as
 \begin{equation}
| \alpha \rangle  \to  \hat \rho (\alpha , \eta , \delta )  \equiv {\frac{2}{\pi\delta}} \int e^{- \frac{2|\beta |^2}{\delta}}  | \sqrt \eta \alpha +\beta \rangle \langle \sqrt \eta \alpha +\beta|  d^2\beta , \label{c-mixture}
\end{equation}
where $\eta$ is the line transmission and $\delta $ is quadrature excess noise. Coherent state is eigen state of $\hat a$: $\hat a | \alpha \rangle = \alpha | \alpha \rangle$.
We define quadrature amplitude $\hat x_1$, $\hat x_2$ by the relation $\hat a = \hat x_1 + i \hat x_2$. $\hat a $ is the annihilation operator of signal pulse mode. The quadrature variance of coherent state is given by $(\Delta x )^2 = \frac{1}{4}$. 
As a frame work, we assume that all noise is caused by Eve in the quantum channel and Bob has an ideal detector. Then, for coherent-state input to the channel (\ref{c-mixture}), Bob observes Gaussian quadrature distribution \cite{hirano} and the observed quadrature variance $(\Delta x_{\textrm{obs}} )^2$ is related to the excess noise as   
 \begin{equation}
(\Delta x_{\textrm{obs}} )^2 = (1+\delta ){(\Delta x )^2}. \label{delta}
\end{equation}
Bob's mean values of quadratures can be related to the transmission and coherent-state amplitude as 
\begin{equation}
\langle \hat x_1 \rangle +i \langle \hat x_2 \rangle = \sqrt \eta \alpha . \label{alpha}
\end{equation}
In terms of $\delta$ and $\eta$, the limitation given by CTA is $\delta < 2 \eta $. We refer to it as classical teleportation limit (CTL) \cite{namiki2}.

\begin{figure}[here]
\begin{center}
\includegraphics[width=8.6cm]{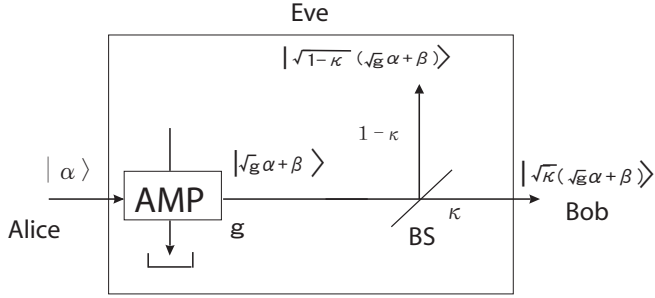}
\caption{\label{amp-bs1} Eve uses a beam splitter (BS) with transmission $\kappa$ after performing a phase-insensitive amplifier (AMP) with amplification gain $g$. $\beta $ is amplification noise.
}
\end{center}
\end{figure}

 Some of eavesdropping attacks which cause the state change (\ref{c-mixture}) can be constructed by combining BSs and phase-insensitive amplifiers (AMP). Simplest case is that Eve uses only one BS and one AMP.
 If Eve uses the Amplifiers after the BS, Eve's and Bob's quadratures are modulated independently. 
 Thus the results of Bob's quadrature measurement and Eve's state are not correlated, and the effectiveness of PS is inherently different from that of BSA \cite{namiki2}. Here we consider the other case which we call amplification-beam-splitting attack (AMPBSA) where Eve inserts BS after AMP (see FIG. \ref{amp-bs1}).


Let us assume that Eve operates a phase-insensitive amplifier \cite{amp} with amplifier gain $g\ge 1$ and inserts a BS with reflectivity $1-\kappa$ (see Fig. \ref{amp-bs1}). Then $g $ and $\kappa$ are related to Bob's mean value and variance of quadratures as
\begin{eqnarray}
\langle \hat x_1 \rangle +i \langle \hat x_2 \rangle &=& \sqrt{ g \kappa}  \alpha , \label{f-gain}\\
(\Delta x_{\textrm{obs}} )^2 &=& \{2(g-1) \kappa +1 \}(\Delta x )^2  \label{kap} .
\end{eqnarray}
Using Eqs. (\ref{delta}), (\ref{alpha}), (\ref{f-gain}) and (\ref{kap}), we obtain
\begin{eqnarray}
g &=& \frac{\eta }{\eta -\delta /2 }, \label{e-gain}\\
\kappa  &=& \eta -\delta /2 .
\end{eqnarray}
From Eq. (\ref{e-gain}), we can see that BSA ($\delta = 0$) is the unit gain case: $g=1$, and CTL ($\delta \to 2 \eta $) is relevant to the infinite-gain limit: $ g \to \infty  $.

Provided Alice sent $|\alpha \rangle$,
Eve's operation makes the joint state of Bob and Eve:
\begin{widetext}
\begin{eqnarray}
| \alpha \rangle _B |0\rangle _E&\to& \hat \rho_{BE} (\alpha,\eta, \delta ) \equiv
\frac{2}{\pi\delta }\int e^{-\frac{2}{\delta }|\beta |^2  }|\sqrt \eta \alpha + \beta  \rangle _{B}\langle \sqrt \eta \alpha + \beta  |  \otimes \left|  {\xi} (\sqrt \eta \alpha + \beta  ) \Big \rangle _{E} \Big\langle  {\xi} (\sqrt \eta \alpha + \beta  )\right| d^2\beta \label{ast},
\end{eqnarray}
\end{widetext}
where we defined 
\begin{eqnarray}
\xi\equiv  \sqrt{\frac{1-\eta +\delta /2}{ \eta -\delta /2 }}.
\end{eqnarray}
The subscripts B and E stand for Bob's and Eve's system, respectively.
We can easily see that $\textrm{Tr}_E \left( \hat \rho _{BE} \right)= \hat \rho (\alpha, \eta, \delta )  $ where $\textrm{Tr}_E $ is partial trace of Eve's system.

The form of $\hat \rho_{BE}$ shows that Eve receives $|  \xi (\sqrt \eta \alpha + \beta  )\rangle $ when Bob receives $|\sqrt \eta \alpha + \beta  \rangle $ i.e., the coherent state Eve receives is different from that of Bob's only by the amplitude factor $ \xi$ for each transmission. This simple picture is useful to explain (i)CTL and (ii)3dB loss limit \cite{coherent} as follows: 

(i) If $\xi$ becomes infinity ($\delta \to 2 \eta $ or $g \to \infty$), Eve can read out the coherent-state amplitude $\sqrt{ \eta } \alpha + \beta $ with arbitrary resolution by performing simultaneous measurement of quadratures. In other words, she can determine the state Bob receives $|\sqrt \eta \alpha + \beta  \rangle $ or she can produce infinite number of copies of this state. This condition is equivalent to the case of intercept-resend attack, and it demonstrates CTL. 

(ii) A sufficient condition for secure key distribution against individual attack is
\begin{equation}I_{AB}\ge I_{AE} \label{cri1},\end{equation}where $I_{AB(AE)}$ is Mutual information between Alice and Bob (Eve) \cite{pa,gpa,postsel}. For Gaussian continuous key distribution protocols \cite{alcon,coherent}, the Mutual information is directly related to the signal-to-noise ratio (SNR); the higher is the SNR, the higher is the amount of mutual information between the parties.
The case $\xi \le 1 $ means Bob's SNR is higher than that of Eve because in Eq. (\ref{ast}) Bob's coherent-state amplitude is larger than that of Eve's. Thus, the security condition given in (\ref{cri1}) is written as
\begin{equation}\xi \le 1  \label{cri2}.\end{equation}
If $\delta = 0$, we obtain a reduced condition $\eta \ge 1/2 $ which gives the 3dB loss limit. This analysis excludes reverse-reconciliation protocol of \cite{coherentR}.

 Now we go into the security of PS protocol against AMPBSA. In the postselection protocol \cite{postsel,hirano,namiki1} , a measurement result $x$ higher than a given threshold $x_0 \ge 0$ is selected to distil the secret key. By setting $x_0$ higher, bit-error rate (BER) of Bob can be arbitrary small provided the quadrature distribution is Gaussian. In contrast to this, if Eve's signal is independent of $x$, Eve's BER remains constant under PS. So PS can make information advantage for Bob compare to Eve. This is not the case for AMPBSA. 
In Eq. (\ref{ast}), 
the modulation $\beta$ is added to both of Eve's and Bob's systems collectively. Through $\beta$, Eve's state depends on $x$ and then Eve's BER depends on PS.
So the correlation between Bob's and Eve's state given by $\beta$ weakens the PS advantage.

Let us consider the case that Alice sends binary-phase-shifted coherent states $| \pm \alpha \rangle $ with $\alpha >0 $ and Bob performs quadrature measurement on the correct basis $\hat x_1$ 
\cite{namiki1,namiki2}. To describe the postselection events, we use Eve's density operator conditioned on Bob's measurement result $x$. 
 If Alice sent $|\alpha \rangle $ and Bob observed $x$, Eve's density operator (conditioned on $x$) is given by
\begin{eqnarray}
\hat \rho _E(\alpha| x ) & = &  \frac{\textrm{Tr}_B \left( \hat \rho _{BE}(\alpha , \eta , \delta) | x\rangle_{B} \langle x | \right) }{P_B(x | \alpha )} \label{bex},
\end{eqnarray} 
where \begin{eqnarray}
{P_B(x | \alpha )}
&=&\textrm{Tr} \left( \hat \rho _{BE}(\alpha, \eta, \delta )  | x\rangle_{B} \langle x | \right)   \nonumber\\
&=&\langle x |\hat \rho (\alpha, \eta, \delta ) | x\rangle 
\end{eqnarray} is the probability that Bob gets quadrature value $x$.

Since Eve does not know the sign of $x$, we can estimate Eve's information from the density operator conditioned on the absolute value of $x$: 
\begin{eqnarray}
\hat \rho _E(\alpha| |x| ) & \equiv &  {P(\alpha |x) \hat \rho _E(\alpha| x ) +P(  \alpha |-x )\hat \rho _E(\alpha| -x )} \label{beax},
\end{eqnarray} where we define
the probability that Alice's choice is $|\alpha \rangle $ when Bob gets $x$: 
\begin{eqnarray}
P(\alpha |x)&\equiv& \frac{P_B(x|\alpha)}{P_B(x|\alpha)+P_B(x|-\alpha)}. 
\end{eqnarray} 

Therefore, if Bob's measurement result is $ \pm x$, Eve gets either of the two mixed states $\hat \rho_E (\pm \alpha ||x| ) $ corresponding to Alice's choice $| \pm \alpha \rangle $, respectively. The next problem is how Eve differentiates the given two signal $\hat \rho_E (\pm \alpha ||x| )$. In general Eve may choose her measurement knowing the value $|x|$.  
Here we restrict our analysis for the case that Eve performs quadrature measurement and determines the bit value according to the sign of her measurement result $x_E$ as Bob does. In this case she does not use the information of $|x|$.

For the PS protocol, Mutual information is written as 
\begin{equation}I_{AB(E)}= \frac{1}{2}\sum_{|x|\ge x_0}P_B(x) i(q_{B(E)}(x)) ,\end{equation}
where 
\begin{equation}P_B(x)=\frac{1}{2}(P_B(x| \alpha)+P_B(x|-\alpha))\end{equation}
is the probability that Bob's measurement result is $x$,  
\begin{equation}i(q)=1+ q \log _2 q + (1-q) \log _2 (1- q)
\end{equation} is Mutual information of binary symmetric channel, and 
$q_{B(E)}(x)$ is BER of Bob (Eve) conditioned on $|x|$.
Since $i(q)$ is a decreasing function of $q$ $(0\le q \le\frac{1}{2})$, $q_B(x) >q_E(x)$ for any $x$ implying that any conditional use of measurement result does not satisfy inequality (\ref{cri1}).

Eve's BER conditioned on $|x|$ is the probability that the signal is $ \hat \rho _E(-\alpha | |x|)$ when the sign of Eve's measurement result $x_E$ is positive:
\begin{widetext}
\begin{eqnarray}
q_{E}(x,\eta ,\delta) &=& \frac{\int _0^\infty \langle x_E | \hat \rho _E (-\alpha | |x|)| x_E \rangle dx_E}{\int _0^\infty \left(  \langle x_E | \hat \rho _E (-\alpha | |x|)| x_E \rangle + \langle x_E | \hat \rho _E (\alpha | |x|)| x_E \rangle \right) dx_E}\\
&=&\frac{ 1}{2}P(\alpha|x)\textrm{ erfc} \left( \sqrt 2\lambda (\delta x +  \sqrt \eta \alpha ) \right)+\frac{ 1}{2}P(-\alpha|x)\textrm{ erfc} \left( \sqrt 2\lambda (\delta x -  \sqrt \eta \alpha ) \right),
\end{eqnarray}
\end{widetext}
where we use the definition of $\hat \rho _{BE}$ (\ref{ast}), Eqs. (\ref{bex} 
- \ref{beax}), and quadrature distribution of coherent state $|\langle x | \alpha \rangle | ^2 =\sqrt{2/ \pi } e^{-2(x- \alpha)^2} $, and we define $\textrm{erfc}(x)=2/\sqrt \pi  \int _s ^\infty e ^{-t^2} dt $ and
\begin{eqnarray}
\lambda &\equiv&\sqrt\frac{(1-\eta )+\delta/2 }{(\eta +\delta/2 )(1+ \delta)}.  \label{efalpha2}
\end{eqnarray}
In what follows we set $x> 0$ for simplicity.

For sufficiently large $x$, $q _E$ is bounded above as 
\begin{widetext}
\begin{eqnarray}
q_{E}(x,\eta ,\delta) &=&\frac{ 1}{2}P(\alpha|x)\textrm{ erfc} \left( \sqrt 2\lambda (\delta x +  \sqrt \eta \alpha )\right)+\frac{ 1}{2}P(-\alpha|x)\textrm{ erfc} \left( \sqrt 2\lambda (\delta x -  \sqrt \eta \alpha ) \right) \nonumber\\
&\le&\frac{ 1}{2}\left\{ \textrm{ erfc} \left( \sqrt 2\lambda (\delta x +  \sqrt \eta \alpha ) \right)+\textrm{ erfc} \left( \sqrt 2\lambda (\delta x -  \sqrt \eta \alpha ) \right) \right\}\nonumber\\
&\le&   \textrm{ erfc} \left( \sqrt 2\lambda (\delta x -  \sqrt \eta \alpha ) \right) \nonumber\\
&\le& \frac{1}{\sqrt \pi} e^{-2\lambda ^2 (\delta x -  \sqrt \eta \alpha )^2}
= \frac{1}{\sqrt \pi} e^{-2\lambda ^2 (\delta^2 x^2 - 2 \sqrt \eta \alpha \delta x +\eta \alpha^2 )}\nonumber\\
&<& \frac{e^{- 2 \lambda ^2 \eta \alpha^2}}{\sqrt \pi} e^{-\lambda ^2 \delta^2 x^2 }  \label{lastineq}.
\end{eqnarray}
\end{widetext}
The first inequality comes from the fact $P(\pm \alpha | x) \le 1$.
The second inequality comes from the fact $\textrm{erfc} (s) $ is a decreasing function of $s$ with $\lambda (\delta x -  \sqrt \eta \alpha )\le  \lambda (\delta x +  \sqrt \eta \alpha )$. Then, we use an inequality $\sqrt \pi \textrm{erfc} (s)  /2= \int _s ^\infty e ^{-t^2} dt \le \int _s ^\infty te ^{-t^2} dt = e^{-s ^2} /2  $ for $s \ge 1$ assuming $x$ is large enough so that $\sqrt 2\lambda (\delta x -  \sqrt \eta \alpha )>1$, which gives the third inequality. Further assuming $x> 4\sqrt \eta \alpha / \delta$, 
we obtain the final expression.

Bob's BER conditioned on $|x|$ is the conditional probability that the sent state is $| - \alpha \rangle$ when his measurement result $x $ is positive: 
\begin{eqnarray}
q_{B}(x,\eta ,\delta)
&\equiv & P(-\alpha |x) \nonumber \\
&=& \frac{1}{ 1+\exp \left[  \frac{8 \sqrt \eta \alpha  x }{  1+ \delta  } \right] } \label{replace}.
\end{eqnarray} 
Since $q_{B} \sim \exp\left( -{\frac{8\sqrt\eta \alpha }{1+ \delta}} x\right)$ and $q_E < e^{-\lambda ^2\delta ^2 x^2} $ from expression (\ref{lastineq}), for any given $\delta > 0$, there exist sufficiently large $x$ where $q_E < q_B$ holds. In such a condition, simple PS setting higher threshold is no more advantageous. It will be more efficient to discard higher quadrature value.

\begin{figure}[htbp]
\begin{center}
\includegraphics[width=8.6cm]{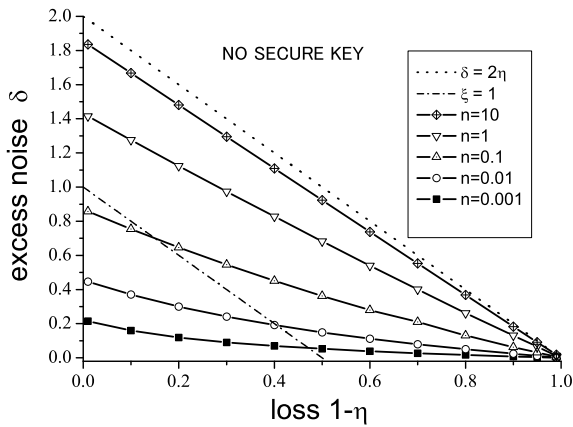}
\end{center}
\caption{ \label{fig1} The security condition $I_{AB}\ge I_E$ cannot be satisfied above the line for mean photon number $n \equiv \alpha ^2 =0.001,\ 0.01,\ 0.1,\ 1,\ 10$ by any postselection of quadrature measurement. For continuous-gaussian-key-distribution protocol the condition can be satisfied below the line $\xi =1 $. $\delta < 2\eta $ is a necessary condition for any coherent-state-QKD protocol given by an intercept-resend attack \cite{namiki2}. 
\label{d-bound}}
\end{figure} 

Figure \ref{d-bound} shows, the parameter region of $\eta $ and $\delta $ which satisfies $q_E \le q_B$ for any choice of $x$ for several mean photon number $n \equiv  \alpha ^2$. This means, above the line, any kind of PS cannot achieve $I_{AB} \ge I_{AE}$. We can see that larger $n$ seems to be more tolearant to noise. This is because, for a given $\delta$,  if $n$ is larger the effect of collective noise is relatively smaller. It should be noted that choice of large $n$ results much smaller secure key gain for higher loss owing to BSA \cite{namiki1}.   
The estimation of secure key gain with some optimization of both Eve's measurement and PS strategy is left for full paper.

In this analysis we assume an ancillary system of amplifier is just traced out (See Fig. 1). It is likely that the ancilla provides some useful information for Eve. In this sense AMPBSA may be weak attack. The condition where Eve cannot access the ancilla can be realized if Alice sends thermal coherent states $ \hat \rho (\alpha , 1, \eta \delta  )$ instead of coherent states and Eve performs just BSA. This case the density operator of joint system is described by Eq. (\ref{ast}) with the replacement $\xi \to \sqrt{(1-\eta)/\eta}$.

In conclusion, we have investigated the security of CV QKD using coherent states against amplification-beam-splitting attack. This attack makes a link between ``direct'' classical-teleportation attack and ``indirect'' beam-splitting attack.
 It has been shown that the postselection protocol setting higher threshold need not ensure the security in the presence of excess Gaussian noise.

We thank M. Koashi for helpful discussions.


\end{document}